\documentclass[11pt]{article}
\usepackage{blois,epsfig}

\def\Journal#1#2#3#4{{#1} {\bf #2}, #3 (#4)}

\def\be{\begin{equation}}
\def\ee{\end{equation}}
\def\bea{\begin{eqnarray}}
\def\eea{\end{eqnarray}}

\newcommand{\kms}{\ensuremath{\hbox{km}\cdot \hbox{s}^{-1}}}
\newcommand{\hmpc}{\ensuremath{h^{-1}\,\hbox{Mpc}}}

\newcommand{\f}{\frac}

\newcommand{\s}{\sigma}

\newcommand{\bfx}{{\bf x}}

\newcommand{\bfy}{{\bf y}}

\newcommand{\bfv}{{\bf v}}

\newcommand{\bfg}{{\bf g}}

\newcommand{\calR}{{\cal R}}
\newcommand{\eps}{{\epsilon}}
\newcommand{\bc}{\begin{center}}
\newcommand{\ec}{\end{center}}

\newcommand{\spose}[1]{\hbox to 0pt{#1\hss}}
\newcommand{\lta}{\mathrel{\spose{\lower 3pt\hbox{$\mathchar"218$}}
 \raise 2.0pt\hbox{$\mathchar"13C$}}}
\newcommand{\gta}{\mathrel{\spose{\lower 3pt\hbox{$\mathchar"218$}}
 \raise 2.0pt\hbox{$\mathchar"13E$}}}

\newcommand{\etal}{{et al.}~}
\newcommand{\err}{r}

\begin{document}

\vspace*{1cm}
\title{PRECISION ANALYSIS OF THE LOCAL GROUP ACCELERATION}

\author{M.J.\ CHODOROWSKI \& P.\ CIECIEL\c AG}
\address{Copernicus Astronomical Center, Bartycka 18, 00-716 Warsaw, 
Poland}

\maketitle\abstracts{
We reexamine likelihood analyses of the Local Group (LG) acceleration,
paying particular attention to nonlinear effects. Under the
approximation that the joint distribution of the LG acceleration and
velocity is Gaussian, two quantities describing nonlinear effects
enter these analyses. The first one is the coherence function,
i.e. the cross-correlation coefficient of the Fourier modes of gravity
and velocity fields. The second one is the ratio of velocity power
spectrum to gravity power spectrum. To date, in all analyses of the LG
acceleration the second quantity was not accounted for. Extending our
previous work, we study both the coherence function and the ratio of
the power spectra. With the aid of numerical simulations we obtain
expressions for the two as functions of wavevector and
$\sigma_8$. Adopting WMAP's best determination of $\sigma_8$, we
estimate the most likely value of the parameter $\beta$ and its
errors. As the observed values of the LG velocity and gravity, we
adopt respectively a CMB-based estimate of the LG velocity, and
Schmoldt et al.'s (1999) estimate of the LG acceleration from the PSCz
catalog. We obtain $\beta = 0.66^{+0.21}_{-0.07}$; thus our errorbars
are significantly smaller than those of Schmoldt et al. This is not
surprising, because the coherence function they used greatly
overestimates actual decoherence between nonlinear gravity and
velocity.}

\section{Introduction}
Comparisons between the CMB dipole and the Local Group (LG)
gravitational acceleration can serve not only as a test for the
kinematic origin of the former but also as a constraint on
cosmological parameters. A commonly applied method of constraining the
parameters by the LG velocity--gravity comparison is a
maximum-likelihood analysis, elaborated by several authors (especially
by Strauss \etal\cite{S92}, hereafter S92). In a
maximum-likelihood analysis, proper objects describing nonlinear
effects are the {\em coherence function} (CF), i.e. the
cross-correlation coefficient of the Fourier modes of the gravity and
velocity fields, and the ratio of the power spectrum of velocity to
the power spectrum of gravity.  Here, with aid of numerical
simulations we model the two quantities as functions of the wavevector
and of cosmological parameters. We then combine these results with
observational estimates of $\bfv_{\rm LG}$ and $\bfg_{\rm LG}$, and
obtain the `best' value of $\beta$ and its errors.

\section{Modelling nonlinear effects}
\label{sec:num}

We follow the evolution of the dark matter distribution using the
pressureless hydrodynamic code CPPA (Cosmological Pressureless
Parabolic Advection, see Kudlicki, Plewa \& R\'o\.zyczka 1996,
Kudlicki \etal 2000 for details). It employs an Eulerian scheme with
third order accuracy in space and second order in time, which assures
low numerical diffusion and an accurate treatment of high density
contrasts. Standard applications of hydrodynamic codes involve a
collisional fluid; however, we implemented a simple flux interchange
procedure to mimic collisionless fluid behaviour.

We studied the CF in an earlier paper.\cite{my} Here we use a
different fitting function, which is more accurate for low values of
$k$:

\begin{equation}
\label{eq:CF_fit}
C(k) = \left [ 1+(a_0 k - a_2 k^{1.5} + a_1 k^2)^{2.5} 
\right ]^{-0.2}.
\end{equation}
Parameters $a_l$ were obtained for 35 different values of $\sigma_8$
in the range [0.1,1], and we found the following, power-law, scaling
relations:
\begin{eqnarray}
a_0 &=& 4.908\ \sigma_8^{0.750} \nonumber \\
a_1 &=& 2.663\ \sigma_8^{0.734} \\
a_2 &=& 5.889\ \sigma_8^{0.714} \nonumber .
\end{eqnarray}
The fit was calculated for $k\in [0, 1]$ $h/\hbox{Mpc}$, with the
imposed constraint $C(k=0)=1$. 

We have found that the ratio of the power spectra, $\calR$,
obtained from simulation can be fitted with the following formula:
\be
\label{eq:pvpgfit_fix}
\calR(k) = [1+(7.071k)^4]^{-\alpha} \,,
\ee
with
\be 
\alpha = -0.06574 + 0.29195\sigma_8 \qquad 
\mathrm{for}~0.3<\sigma_8<1 \,.
\ee

\section{Parameter estimation}
\label{sec:param}
We apply our formalism to the PSCz survey. As the value of $\s_8$ we
adopt its WMAP's estimate, $\s_8 = 0.84$ ($\pm 0.04$; Spergel et
al. 2003). This specifies the coherence function and the ratio of the
power spectrum of velocity to the power spectrum of gravity. Also,
this provides a normalization for the power spectrum of density.

\subsection{Parameter dependence of the model}
The likelihood of specific values of $\beta$ and of the linear bias,
$b$, is determined by the following distribution (Juszkiewicz \etal
1990, Lahav, Kaiser \& Hoffman 1990):

\be
f(\bfg,\bfv) = \f{(1 - \err^2)^{-3/2}}{(2 \pi)^{3} \s_\bfg^{3} 
\s_\bfv^{3}} 
\exp\left[- \f{x^2 + y^2 - 2 \err \cos\psi\,x y}{2(1 - \err^2)}\right] 
\,,
\label{eq:dist}
\ee

where $\s_\bfg$ and $\s_\bfv$ are respectively the r.m.s.\ values of a
single Cartesian component of gravity and velocity, $(\bfx,\bfy) =
(\bfg/\s_\bfg,\bfv/\s_\bfv)$, and $\psi$ is the misalignment angle
between $\bfg$ and $\bfv$. Finally, $\err$ is the cross-correlation
coefficient of $g_m$ with $v_m$, where $g_m$ ($v_m$) denotes an
arbitrary Cartesian component of $\bfg$ ($\bfv$).

In this distribution, the observables are $\bfg$ and $\bfv$, or $g$,
$v$, and the misalignment angle, $\psi$. Following Schmoldt
\etal\cite{S99} (hereafter S99), we adopt for them the following
values: $g = 933 \;\kms$ (from the distribution of the PSCz galaxies
up to $150\;\hmpc$), $v = 627\; \kms$ (inferred from the 4-year COBE
data by Lineweaver \etal 1996), and $\psi = 15^\circ$.

\begin{figure}
\centerline{\includegraphics[angle=0,scale=0.5]{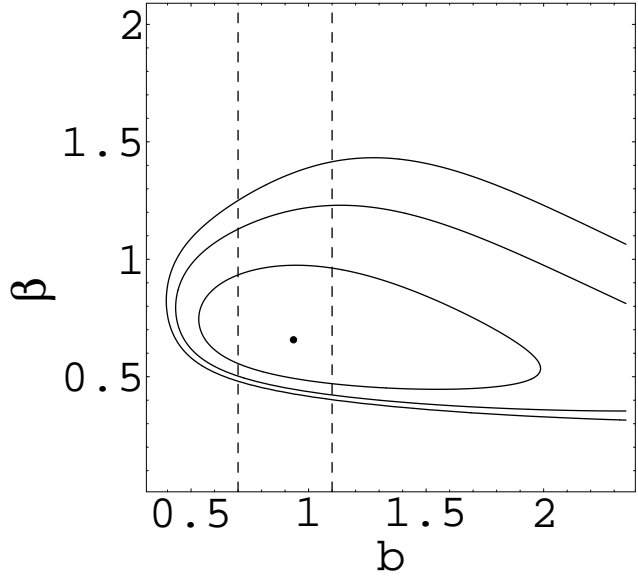}}
\caption{\label{fig:lh} Likelihood contours for the parameters $\beta$
and $b$, corresponding to the confidence levels of 68, 90 and
95\%. The maximum of the likelihood function is denoted by a dot.}
\end{figure}

The theoretical quantities are $\s_\bfg$, $\s_\bfv$, and $\err$. The
variance of a single spatial component of measured gravity,
$\s_\bfg^2$, is a sum of the cosmological component, $\s_{\bfg,c}^2$,
and errors, $\eps^2$. Since gravity here is inferred from a galaxian,
rather than mass, density field, we have $\s_{\bfg,c}^2 = b^2
s_\bfg^2$, where $s_\bfg^2$ is the variance of a single component of
the true (i.e., mass-induced) gravity,
\begin{equation}
s_\bfg^2 = \frac{1}{6\pi^2}\int_0^{\infty} \widehat{W}_g^2(k) P(k) dk
\,.
\label{eq:s_g}
\end{equation}

The gravity errors are twofold: due to finite sampling of the galaxy
density field, and due to the reconstruction of the galaxy density
field in real space. Therefore, $\eps^2 = (\s_{\rm SN}^2 + \s_{\rm
rec}^2)/3$, where $\s_{\rm SN}^2$ and $\s_{\rm rec}^2$ are
respectively the shot noise (or, sampling) variance and the
reconstruction variance. In brief,
\be
\s_\bfg^2 = b^2 s_\bfg^2 + \f{\s_{\rm SN}^2 + \s_{\rm rec}^2}{3} \,,
\label{eq:sigma_g}
\ee
where $\s_{\rm SN} = 160\; \kms$, and $\s_{\rm rec} = 58\; \kms$
(S99). Next, we have
\be
\s_\bfv = \frac{\Omega_m^{0.6}}{6\pi^2}\int_0^{\infty} 
\widehat{W}_v^2(k) \calR(k) P(k) dk \,.
\label{eq:s_v}
\end{equation}
Finally, errors in the estimate of the LG gravity do not affect the
cross-correlation between the LG gravity and velocity, but increase
the gravity variance. This has the effect of lowering the value of the
cross-correlation coefficient. Specifically, we have
\be
\err = \rho \left(1+ \f{\s_{\rm SN}^2 + 
\s_{\rm rec}^2}{3 b^2 s_\bfg^2} \right)^{-1/2} ,
\label{eq:err_errors} 
\ee 
where
\be
\rho = \f{\int_0^\infty \widehat{W}_\bfg(k) \widehat{W}_\bfv(k)
{C}(k) \calR^{1/2}(k) P(k) dk}{\left[\int_0^\infty 
\widehat{W}_\bfg^2(k) P(k) dk\right]^{1/2} \left[\int_0^\infty 
\widehat{W}_\bfv^2(k) \calR(k) P(k) dk\right]^{1/2}} \,.
\label{eq:scaled_err} 
\ee 
Thus, the likelihood depends explicitly on the parameters $b$ and
$\Omega_m$, or on $b$ and $\beta \equiv \Omega_m^{0.6}/b$. 

\subsection{Joint likelihood for $\beta$ and b}

Figure~\ref{fig:lh} shows isocontours of the joint likelihood for
$\beta$ and $b$, corresponding to the confidence levels of 68, 90 and
95\%. The maximum of the likelihood is denoted by a dot. The
corresponding values of $\beta$ and $b$ are respectively $0.66$ and
$0.94$. The isocontours of the likelihood are much more elongated
along the $b$-axis than along the $\beta$-axis, making the resulting
constraints on $\Omega_m$ much weaker than on $\beta$. In practice,
therefore, from our analysis we cannot say much about bias, or
$\Omega_m$, alone. On the other hand, we can put fairly tight
constraints on $\beta$.

\begin{figure}
\centerline{\includegraphics[angle=0,scale=0.5]{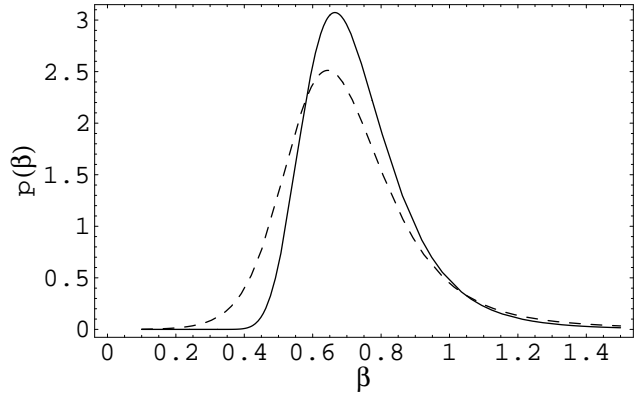}}
\caption{\label{fig:lh_marg} The marginal distribution for $\beta$. 
The result of marginalizing over all possible values of $b$ (from zero
to infinity) is shown as a dashed line. The result of marginalizing
over the values of $b$ in the range $[0.7, 1.1]$ is shown as a solid
line.}
\end{figure}

To do this, we use the fact that the square root of the variance of
the PSCz galaxy counts at 8 \hmpc\ is $\s_8^{PSCz} \simeq 0.75$
(Sutherland et al. 1999). Combined with WMAP's estimate of $\s_8$,
this yields for the bias of the PSCz galaxies the value about
$0.9$. Therefore, we adopt here a conservative prior for the bias,
namely that it is constrained to lie in the range $[0.7, 1.1]$. These
limits are marked in Figure~\ref{fig:lh} as dashed vertical lines. We
marginalize the likelihood over the values of $b$ in this range. The
resulting distribution for $\beta$ is shown in
Figure~\ref{fig:lh_marg} as a solid line. We obtain $\beta =
0.66^{+0.21}_{-0.07}$ ($68$\% confidence limits).

\section{Summary and conclusions}
\label{sec:conc}

The analysis of the LG acceleration performed by S92 and S99 was in a
sense more sophisticated than ours. Both teams analysed a differential
growth of the gravity dipole in subsequent shells around the
LG. Instead, here we used just one measurement of the total
(integrated) gravity within a radius of $150\;\hmpc$. Nevertheless,
the errors on $\beta$ we have obtained are significantly smaller than
those of S92 and S99. In particular, S99 obtained $\beta =
0.70^{+0.35}_{-0.20}$ at $1\s$ confidence level. It is striking that
while our best value of $\beta$ is close to theirs, our errors are
significantly smaller. The reason is our careful modelling of
nonlinear effects. In a previous paper we showed that the coherence
function used by S99 greatly overestimates actual decoherence between
nonlinear gravity and velocity. Tighter correlation between the LG
gravity and velocity should result in a smaller random error of
$\beta$; in the present work we have shown this to be indeed the case.

\section*{Acknowledgments}
This research has been supported in part by the Polish State Committee
for Scientific Research grants No.~2.P03D.014.19 and 2.P03D.017.19.

\end{document}